\documentclass[12pt] {article}
\usepackage{latexsym}{\rm }
\usepackage{graphics}
\usepackage{graphpap}
\usepackage{color}

\newcommand{\beq}{\begin{equation}}
\newcommand{\feq}[1]{\label{#1} \end{equation}}
\newcommand{\beqr}{\begin{eqnarray}}
\newcommand{\feqr}{\end{eqnarray}}
\def\non{\nonumber}
\def\noi{\noindent}
\def\slasha#1{\setbox0=\hbox{$#1$}#1\hskip-\wd0\hbox to\wd0{\hss\sl/\/\hss}}
\def\slashb#1{\setbox0=\hbox{$#1$}#1\hskip-\wd0\dimen0=5pt\advance
       \dimen0 by-\ht0\advance\dimen0 by\dp0\lower0.5\dimen0\hbox
         to\wd0{\hss\sl/\/\hss}}

\newcommand{\rf}[1]{(\ref{#1})}

\makeatletter
\newcommand{\sign}{\mathop{\operator@font sign}}
\makeatother
\makeatletter
\newcommand{\const}{\mathop{\operator@font const}}
\makeatother
\def\np#1#2#3{Nucl. Phys. {\bf{B#1}} (#2) #3}

\def\rmp#1#2#3{Rev. Mod. Phys. {\bf{#1}} (#2) #3}

\def\jmp#1#2#3{J. Math. Phys. {\bf{#1}} (#2) #3}

\def\jhp#1#2#3{JHEP {\bf{#1}} (#2) #3}
\renewcommand{\thefootnote}{\fnsymbol{footnote}}

\begin{document}

\begin{center}


{\Large \bf Noncommutative Quantization in 2D Conformal Field Theory}\\ [4mm]

\large{Agapitos Hatzinikitas}\footnote [2] 
{Email: ahatzini@tem.uoc.gr} \\ [5mm]

{\small University of Crete, \\
Department of Applied Mathematics, \\
L. Knosou-Ambelokipi, 71409 Iraklio Crete,\\ 
Greece}\\ [5mm]

\large{and} \\ [5mm]

\large{Ioannis Smyrnakis} \footnote [3] 
{Email: smyrnaki@tem.uoc.gr}\\ [5mm]

{\small University of Crete, \\
Department of Applied Mathematics, \\
L. Knosou-Ambelokipi, 71409 Iraklio Crete,\\ 
Greece}
\vspace{5mm}

\end{center}

\begin{abstract}
The simplest possible noncommutative harmonic oscillator in two dimensions is used to quantize the free closed
bosonic string in two flat dimensions.  The partition function is not deformed by the 
introduction of noncommutativity, if we rescale the time and change the compactification radius appropriately. 
The four point function is deformed, 
preserving, nevertheless, the sl(2,C) invariance.  
Finally the first Ward identity of the deformed theory is derived.

\end{abstract}
\newpage

\renewcommand{\thefootnote}{\arabic{footnote}}
\setcounter{footnote}{0}


In this letter we explore a possible way of quantizing a string based on a noncommutative harmonic oscillator 
as opposed to the ordinary commutative one \cite{joel}.  This noncommutativity is neither the one in 
the D brane worldvolume that arises in the  quantization of open strings ending on D 
branes with background B field \cite{chu}, nor the one introduced in M-theory compactified on $T^2$ \cite{douglas}.   
\par To study the modifications that occur in bosonic string theory, the simplest possible 
noncommutative harmonic oscillator in two dimensions is employed.  We assume that the position and momentum operators 
satisfy the following commutation relations \cite{aj}:
\beqr
[\hat{q}_n^i,\hat{p}_n^j]=i\hbar \delta_{ij}; \quad [\hat{q}_n^1,\hat{q}_n^2]=i\theta/n; \quad 
[\hat{p}_n^1,\hat{p}_n^2]=-in\theta .
\label{oscill}
\feqr
\noi An operator representation of the $\hat{q}_n^i$($\hat{p}_n^i$) that realizes the commutation relations \rf{oscill} is: 
\beqr
\hat{q}_n^1 &=& \sqrt{\frac{\alpha }{2n}}(a_n^1+{a_n^1}^\dagger )\\ 
\hat{p}_n^1 &=& -i\sqrt{\frac{n}{2\alpha}}\bigl[(\hbar a_n^1-i\theta a_n^2)-(\hbar {a_n^1}^\dagger +i\theta {a_n^2}^\dagger ) \bigr] \\
\hat{q}_n^2 &=& \sqrt{\frac{1}{2n\alpha}}\bigl[(\hbar a_n^2-i\theta a_n^1)+(\hbar {a_n^2}^\dagger +i\theta {a_n^1}^\dagger ) \bigr]
\equiv \sqrt{\frac{\alpha }{2n}}(A_n+A_n^\dagger )\\
\hat{p}_n^2 &=& -i\sqrt{\frac{n\alpha }{2}}(a_n^2-{a_n^2}^\dagger),
\label{repr}
\feqr
where $\alpha =\sqrt{\hbar^2+\theta^2}$. The creation and annihilation 
operators satisfy the usual commutation relations: $[a_n^i,{a_n^j}^\dagger]=\delta_{ij}$, from which, one deduces 
that $[A_n,A_n^\dagger ]=1$. 
The Hamiltonian of this harmonic oscillator is $H=\alpha \sum_{i=1}^2 n {a_n^i}^\dagger a_n^i+ \const \!.$  
and the time evolution of the modes 
$a_n^i,{a_n^i}^\dagger$ is found to be $a_n^i(t)=a_n^ie^{-i\frac{a}{\hbar}nt}$ and  
${a_n^i}^\dagger (t)={a_n^i}^\dagger e^{i\frac{a}{\hbar}nt}$.

\par Consider now the free closed bosonic string in two dimensions.  The coordinates that 
satisfy both the equations of motion and the boundary conditions admit a real expansion of the form 
\beqr
\tilde{X^1}(\sigma,t) =X^1_0 + \frac{P^1_0}{2}t + \frac{N^1_0}{2}\sigma + 
\sum_{n>0}q_n^1(t)\cos n\sigma +\bar{q}_n^1(t)\sin n\sigma \\
\tilde{X^2}(\sigma,t) =X^2_0 + \frac{P^2_0}{2}t + \frac{N^2_0}{2}\sigma +
\sum_{n>0}q_n^2(t) \cos n\sigma +\bar{q}_n^2(t)\sin n\sigma.
\label{expn1}
\feqr
\noi We treat $q_n^1,q_n^2$ and  $\bar{q}_n^1,\bar{q}_n^2$  as two separate systems of two dimensional 
harmonic oscillators. Upon quantization the operators $\hat{q}_n^1,\hat{q}_n^2$ and  
$\hat{\bar{q}}_n^1,\hat{\bar{q}}_n^2$ are expressed in terms of the operators $a_n,a_n^\dagger $ and $b_n,b_n^\dagger $ 
respectively,
obeying identical commutation relations \rf{oscill}.  Rescaling the modes by $a_n^i\rightarrow a_n^i/\sqrt{n}$, 
$b_n^i\rightarrow b_n^i/\sqrt{n}$ and defining $a_{-n}^1=-{a_n^1}^\dagger$, $a_{-n}^2={a_n^2}^\dagger$, 
$b_{-n}^1={b_n^1}^\dagger$, $b_{-n}^2=-{b_n^2}^\dagger$ we have
\beqr
\tilde{X^1}(\sigma,t)=X^1_0 + \frac{P^1_0}{2}t + \frac{N^1_0}{2}\sigma + 
\frac{i}{2}\sum_{n \neq 0}\frac{1}{n} \left(C_ne^{-in(\frac{a}{\hbar}t-\sigma)}+
\bar{C_n}e^{-in(\frac{a}{\hbar}t+\sigma)} \right) \\
\tilde{X^2}(\sigma,t)=X^2_0 + \frac{P^2_0}{2}t + \frac{N^2_0}{2}\sigma +
\frac{i}{2}\sum_{n \neq 0}\frac{1}{n} \left(D_ne^{-in(\frac{a}{\hbar}t-\sigma)}+
\bar{D_n}e^{-in(\frac{a}{\hbar}t+\sigma)} \right)
\label{expn2}
\feqr
\noi where 
\beqr
C_n = -\sqrt{\frac{\alpha}{2}} \left(b_n^1+i a_n^1 \right); &\quad&
\bar{C}_n=\sqrt{\frac{\alpha}{2}} \left(b_n^1-i a_n^1 \right) \\
D_n = -\sqrt{\frac{\alpha}{2}} \sign (n) \left(B_n + i A_n \right); &\quad&
\bar{D}_n=\sqrt{\frac{\alpha}{2}} \sign (n) \left(B_n-i A_n \right) \\
A_n =  \frac{\hbar a_n^2 -i\theta a_n^1}{\alpha }; &\quad &
B_n= \frac{\hbar b_n^2 -i\theta b_n^1}{\alpha }.
\label{def1}
\feqr
\noi The C and D modes obey the following commutation relations:
\beqr
[C_n,C_{-n}]=n\alpha; \quad [C_n,D_{-n}]=i n \theta  \sign (n); \quad  [D_n,D_{-n}]=n \alpha 
\label{com1}
\feqr
\noi where the corresponding barred operators satisfy the same commutation relations and commute with the 
unbarred operators. The zero modes satisfy the commutation relations
\beqr
\label{zerocom}
[X^i_L,P^j_L]=[X^i_R,P^j_R]=i\alpha \delta_{ij}
\feqr
where $X^i_0=(X^i_L + X^i_R)/2$ and
\beqr
\label{defzer}
P^i_L=\frac{\hbar}{\alpha}\frac{P^i_0}{2}+\frac{N^i_0}{2}; \qquad P^i_R=\frac{\hbar}{\alpha}\frac{P^i_0}{2}-\frac{N^i_0}{2}. 
\feqr
The reason for adopting the commutation relations \rf{zerocom} will be justified when we consider propagators in Euclidean 
space-time \rf{propag}.
\par The regularized Hamiltonian, which is the sum of the Hamiltonians of the individual harmonic oscillators, is: 
\beqr
H&=&\frac{1}{2}\sum_{i=1}^2\left(\left(P^i_L\right)^2 + \left(P^i_R\right)^2 \right)
+ \sum_{n>0}\Bigg[\frac{\alpha^2}{\hbar^2}(C_{-n}C_n+\bar{C}_{-n}\bar{C}_n+D_{-n}D_n+\bar{D}_{-n}\bar{D}_n)
\non \\
&+& i\frac{\alpha \theta}{\hbar^2}(D_{-n}C_n-C_{-n}D_n+\bar{D}_{-n}\bar{C}_n-\bar{C}_{-n}\bar{D}_n)\Bigg]+ 
\frac{\alpha}{12}.
\label{hamilt2}
\feqr
\noi The term $\alpha/12$ stems from the normal ordering. This is the generator of time translations on the 
coordinates $X^i$.  The generator of spatial translations is:
\beqr
P&=&\frac{\hbar}{2\alpha} \sum_{i=1}^2 \left( \left(P^i_R\right)^2 - \left(P^i_L\right)^2 \right)
+\sum_{n>0}\Bigg[\frac{\alpha}{\hbar}(C_{-n}C_n-\bar{C}_{-n}\bar{C}_n+D_{-n}D_n-\bar{D}_{-n}\bar{D}_n)\non \\
&+& i\frac{\theta}{\hbar}(D_{-n}C_n-C_{-n}D_n-\bar{D}_{-n}\bar{C}_n+\bar{C}_{-n}\bar{D}_n)\Bigg].
\label{mom1}  
\feqr
\noi Note that the momentum and the Hamiltonian operator commute, as expected.  
\par Performing the Wick rotation and passing to the complex plane 
($t \rightarrow -it$, $\theta\rightarrow -i\theta$, $\tau=-i\frac{\alpha }{\hbar }t$, $w=\tau+i\sigma $, $z=e^{-w}$),
we obtain
\beqr
X^1(z)=x_R^1-ip_R^1 \ln z+i\sum_{n\ne 0}\frac{1}{n}C_nz^{-n} \\
\bar{X}^1(\bar{z})=x_L^1-ip_L^1 \ln \bar{z}+i\sum_{n\ne 0}\frac{1}{n}\bar{C}_n\bar{z}^{-n} \\
X^2(z)=x_R^2-ip_R^2 \ln z+i\sum_{n\ne 0}\frac{1}{n}D_nz^{-n} \\
\bar{X}^2(\bar{z})=x_L^2-ip_L^2 \ln \bar{z}+i\sum_{n\ne 0}\frac{1}{n}\bar{D}_n\bar{z}^{-n}
\label{expn3}
\feqr
\noi where $X^i(\sigma ,\tau )=(X^i(z)+\bar{X}^i(\bar{z}))/2$.
\par  The partition function on the torus for this prototype theory is given by: 
\beqr
Z=Tr(e^{2\pi i\frac{\tau_1}{\hbar}P}e^{-2\pi \frac{\hbar}{\alpha }\frac{\tau_2}{\hbar}H})
\label{part1}
\feqr
\noi where the trace is taken over the module generated by the negative modes of C and D operators 
subjected to the commutation relations \rf{zerocom}.  After evaluating this trace we get 
the usual expression as if noncommutativity was absent:
\beqr
Z=\frac{Z_0^2(\tau, \bar{\tau})}{|\eta(\tau)|^4}
\label{part2}
\feqr
where 
\beqr
\label{defZ0}
Z_0 (\tau, \bar{\tau})=\sum_{n,m=-\infty}^{\infty}
q^{\frac{1}{2}\left(\sqrt{\alpha}\frac{m}{2R}-\frac{nR}{\sqrt{\alpha}} \right)^2}
\bar{q}^{\frac{1}{2}\left(\sqrt{\alpha}\frac{m}{2R}+\frac{nR}{\sqrt{\alpha}} \right)^2},
\feqr
$q=e^{2i\pi \tau}$ and $R$ is the compactification radius. The quantization of the zero modes takes the form
\beqr
\label{quaze}
P^i_0=\frac{\alpha^2}{\hbar}\frac{m}{R};  \qquad N^i_0=2\pi n R .
\feqr
\par To make further progress we need to determine the propagators.  These turn out to be: 
\beqr
\label{propag}
<X^1(z)X^1(w)>=<(x_R^1)^2>-\alpha \ln(z-w)  \non \\
<X^2(z)X^2(w)>=<(x_R^2)^2>-\alpha \ln(z-w) \non \\
<X^1(z)X^2(w)>=<x_R^1x_R^2>-\theta \ln(1-\frac{w}{z}) \non \\
<X^2(z)X^1(w)>=<x_R^2x_R^1>+\theta \ln(1-\frac{w}{z}). 
\feqr
\noi In determining \rf{propag} we have imposed the commutation relations \rf{zerocom} among the zero modes 
which guarantee the $\ln(z-w)$ dependence of $<X^1(z)X^1(w)>$ and $<X^2(z)X^2(w)>$ .
Note that there is a new singularity introduced at $z=0$, but it is not possible to avoid it by changing the zero 
mode commutation relations because then it becomes impossible to find generators of time and space translations. 
Identical relations hold in the antiholomorphic sector.  
\par Regarding the stress-energy tensor, there is no longer a unique generator of conformal transformations 
for both $X^i$ components.  Rather there is the usual tensor $T_1(z)=-\frac{1}{2\alpha}:(\partial_zX^1(z))^2:$ that 
generates conformal transformations on $X^1(z)$ but not on $X^2(z)$ and conversely for $T_2(z)$. 
The algebra of the moments of each 
stress-energy tensor is the Virasoro algebra with central charge one. Nevertheless the moments of $T_1$, $T_2$ do not 
commute. The primary fields are the usual ones for 
each string component.  
\par Our next task is to compute the correlation functions on the sphere.  The only interesting two point function
is the mixed one:
\beqr
<0|:e^{ik_1X^1(z)}::e^{ik_2X^2(w)}:|0>=<0|:e^{ik_1x_R^1}::e^{ik_2x_R^2}:|0>(1-\frac{w}{z})^{\theta k_1k_2}=0.
\label{corr1}
\feqr
\noi It vanishes because of the expectation value of the zero modes, unless both $k_1,k_2$ are 0.  The zero 
mode expectation value also indicates that in higher correlation functions charge conservation must be 
maintained for each primary field separately.  This implies that the first correlation function that will 
differ from the commutative case is the four point function.  It takes the form: 
\beqr
<:e^{ikX^1(u)}::e^{-ikX^1(v)}::e^{i\lambda X^2(w)}::e^{-i\lambda X^2(z)}:>=
(u-v)^{-\alpha k^2}(w-z)^{-\alpha \lambda^2}\Bigg[ \frac{(u-w)(v-z)}{(u-z)(v-w)} \Bigg]^{\theta k\lambda }.
\label{corr2}
\feqr
\noi This four point function is invariant under global sl(2,C) transformations, a fact that indicates the 
dependence of the  correlation functions on the complex structure only.  Thus the noncommutative quantization procedure 
we apply gives a nontrivial deformation of the original theory, since the four point function is 
deformed, while the nice geometric properties of the four point functions are preserved.  
\par  We complete our discussion by writing down the first 
Ward identity.  Using the operator product expansion of the energy momentum tensors we get the formula 
\beqr
&&\sum_{i=1}^{N}<:e^{ik_1 X^{n_1}(w_1)}: \cdots \frac{1}{2\pi i} \oint_{w_i} \epsilon (z) T_{n_i}(z) 
:e^{ik_i X^{n_i}(w_i)}: \cdots :e^{ik_N X^{n_N}(w_N)}:> \non \\
&=& \sum_{i=1}^{N} \frac{1}{2\pi i} \oint_{w_i} \Bigg(\frac{\alpha k_i^2}{2} \frac{1}{(z-w_i)^2} 
+ \frac{1}{z-w_i} \partial_{w_i} \Bigg)\epsilon (z) 
<:e^{ik_1 X^{n_1}(w_1)}: \cdots :e^{ik_N X^{n_N}(w_N)}:>
\label{ward1}
\feqr
which takes the form
\beqr
&&\frac{1}{2\pi i}\int_{C_1}\epsilon(z)<T_1 (z):e^{ik_1 X^{n_1}(w_1)}: \cdots :e^{ik_N X^{n_N}(w_N)}:> \non \\
&+&
\frac{1}{2\pi i}\int_{C_2}\epsilon(z)<T_2 (z):e^{ik_1 X^{n_1}(w_1)}: \cdots :e^{ik_N X^{n_N}(w_N)}:> \non \\
&=& \frac{1}{2\pi i} \int_{C}\epsilon(z)\sum_{i=1}^{N} \Bigg(\frac{\alpha k_i^2}{2} \frac{1}{(z-w_i)^2} 
+ \frac{1}{z-w_i} \partial_{w_i} \Bigg) <:e^{ik_1 X^{n_1}(w_1)}: \cdots :e^{ik_N X^{n_N}(w_N)}:>.
\label{ward2}
\feqr
\noi The contour $C_1$ encircles the points $w_i$ that appear inside $e^{ik_i X^1(w_i)}$ while the contour $C_2$ 
encircles the points $w_i$ that appear inside $e^{ik_i X^2(w_i)}$. The contour $C$ encloses all insertion points.
\par As a conclusion, we have shown that it is possible to deform nontrivially the closed bosonic string in two 
flat dimensions by following 
a quantization procedure based on a noncommutative harmonic oscillator. This deformation is shown to preserve sl(2,C) 
invariance of the four point function. Quantization using a more general noncommutative harmonic oscillator 
is currently under investigation.  

\color{black}



\bibliographystyle{plain}

\begin{thebibliography} {99}

\bibitem{joel} J. Scherk, ``\textit{An Introduction to the Theory of Dual Models and Strings}", \rmp{47}{1975}{123};\\
M.B. Green, J. H. Schwarz and E. Witten, ``\textit{Superstring Theory}", Cambridge University Press, 1987 ; \\
P. Ginsparg, ``\textit{Applied Conformal Field Theory}", in  ``\textit{Les Houches, 1988}", edited by E. Br$\acute{e}$zin 
and J. Zinn-Justin, Elsevier Publishers, Amsterdam, 1989.
    

\bibitem{chu} C.-S. Chu and P.-M. Ho, ``\textit{Noncommutative Open String and D-brane}", \np{550}{1999}{151}.

\bibitem{douglas} A. Connes, M. R. Douglas and A. Schwarz, ``\textit{Noncommutative Geometry and Matrix Theory:
Compactification on Tori}", \jhp{02}{1998}{003}.


\bibitem{aj} A. Hatzinikitas and I. Smyrnakis, ``\textit{The Noncommutative Harmonic Oscillator in More Than One Dimension}'', 
\jmp{43}{2002}{113}, and references therein.

\end{thebibliography}


\end{document}